\documentclass[12pt]{article}

\usepackage[square,comma,sort&compress]{natbib}
\usepackage{graphicx}
\usepackage{axodraw}
\usepackage[dvips]{color}
\usepackage{epsfig}
\def \nbb {$\beta\beta_{0\nu}$ }
\def\e6{$E(6)$}
\def\10{$SO(10)$}
\def\21{$SU(2) \otimes U(1) $}

\def\422{$SU(4) \otimes SU(2) \otimes SU(2)$}
\def\321{$SU(3) \otimes SU(2) \otimes U(1)$}
\def\lsim{\raise0.3ex\hbox{$\;<$\kern-0.75em\raise-1.1ex\hbox{$\sim\;$}}}
\def\gsim{\raise0.3ex\hbox{$\;>$\kern-0.75em\raise-1.1ex\hbox{$\sim\;$}}}
\let\vev\VEV

\DeclareMathAlphabet{\mathsc}{OT1}{cmr}{m}{sc}
\newcommand{\ed}{\end{document}}

\newcommand{\Atm}  {\mathsc{atm}}

\newcommand{\Dma}  {\Delta m^2_\Atm}
\newcommand{\AHEP}{AHEP Group, Instituto de F\'{\i}sica Corpuscular --
  C.S.I.C./Universitat de Val{\`e}ncia \\
  Edificio de Institutos de Paterna, Apartado 22085,
  E--46071 Val{\`e}ncia, Spain\\}

\begin{document}
\begin{titlepage} 

\begin{flushright}
IFIC/03-56\\
\end{flushright} 
\vspace*{3mm} 

\begin{center}  

\textbf{\large Phenomenological Tests of Supersymmetric  
   \\ $A_4$ Family Symmetry Model of Neutrino Mass}\\[10mm]

{M. Hirsch${}^1$,  J. C. Rom\~ao${}^{2}$, 
S. Skadhauge${}^2$, J. W. F. Valle${}^1$ and  A. Villanova del Moral${}^1$} 
\vspace{0.3cm}\\ 

{\it $^1$ \AHEP}

{$^2$ Departamento de F\'\i sica and CFIF, Instituto Superior T\'ecnico\\
  Av. Rovisco Pais 1, $\:\:$ 1049-001 Lisboa, Portugal \\}

\end{center}

\vspace*{3mm}

\begin{abstract}
  
  Recently Babu, Ma and Valle proposed a model of quark and lepton
  mixing based on $A_4$ symmetry\cite{babu:2002dz}.  Within this model
  the lepton and slepton mixings are intimately related. We perform a
  numerical study in order to derive the slepton masses and mixings in
  agreement with present data from neutrino physics.  We show that,
  starting from three-fold degeneracy of the neutrino masses at a high
  energy scale, a viable low energy neutrino mass matrix can indeed be
  obtained in agreement with constraints on lepton flavour violating
  $\mu$ and $\tau$ decays.  The resulting slepton spectrum must
  necessarily include at least one mass below 200 GeV which can be
  produced at the LHC.  The predictions for the absolute Majorana
  neutrino mass scale $m_0 \geq 0.3$ eV ensure that the model will be
  tested by future cosmological tests and \nbb searches. Rates for
  lepton flavour violating processes $\ell_j \to \ell_i + \gamma$ in
  the range of sensitivity of current experiments are typical in the
  model, with BR$(\mu \to e \gamma) \gsim 10^{-15}$ and the lower
  bound BR$(\tau \to \mu \gamma) > 10^{-9}$.  To first approximation,
  the model leads to maximal leptonic CP violation in neutrino
  oscillations.

\end{abstract} 
 
\end{titlepage} 

\newpage

\setcounter{page}{1} 

\section{Introduction}

The remarkable experimental achievements in neutrino
physics~\cite{fukuda:1999pp,fukuda:1998ah,fukuda:1998mi,fukuda:2001nk,
  ahmad:2002jz,ahn:2002up,apollonio:1999ae,Ahmed:2003kj} have provided
great insight in the neutrino masses and mixings. In particular it is
now well established that the leptonic mixing matrix is rather
different from the quark mixing matrix~\cite{maltoni:2003da}. The
structure of the mixings suggested by experiment involves a large
mixing angle describing solar neutrino oscillations, a maximal one
describing atmospheric neutrino oscillations, and a small one to
account for reactor neutrino data. This is in sharp contrast to the
three small mixings that characterize the quark sector and poses a
challenge to models of the origin of the flavour structure.

In this paper we perform a detailed study of the model put forward by
Babu, Ma and Valle in Ref.\cite{babu:2002dz}.  The model offers a
simple and coherent picture of the quark and lepton mixings. Both
mixing matrices are generated by radiative corrections, but with
different tree-level structures fixed at some high energy scale, which
we will denote by $M_N$. Small off-diagonal corrections to a
hierarchical mass matrix will give small mixing angles. In contrast
large mixing is a natural consequence of small corrections to
degenerate energy levels. Therefore, the quark mixings are pushed to
be small due to the hierarchical structure of their masses.  Whereas,
the large solar mixing angle is achieved due to degeneracy of the
neutrino masses at tree-level. The two other leptonic mixing angles
are fixed by the family symmetry.

The model uses $A_4$ family symmetry, where $A_4$ is the symmetry
group of the tetrahedron or equivalently the group of even
permutations of four elements. The family symmetry is broken at the
high energy scale $M_N$, which is imagined to be around the scale of
grand unification (of order $10^{16}$ GeV).  However, the model is not
explicitly embedded into any grand unification group. In order to have
a natural stabilization of the different energy scales involved, low
energy supersymmetry (SUSY) is used.  Besides, as will be discussed
below, the soft SUSY breaking terms constitute a necessary ingredient
of the model, implying sizeable flavour changing interactions.

In addition to the usual fields in the Minimal Supersymmetric Standard
Model (MSSM) a number of heavy fermion and Higgs fields are
introduced.  Within the $A_4$ family symmetry scheme, this implies: \\
(i) the quark mass matrices are hierarchical and aligned, hence giving
$V_{\rm CKM}(M_N)=I$. As a result the low energy CKM
angles are naturally small;\\
(ii) all three neutrino masses are exactly degenerate at $M_N$, with
an off-diagonal $\nu_\mu - \nu_\tau$ texture.  The atmospheric mixing
angle is thereby predicted to be maximal and this feature is
kept even after the leading radiative corrections;  \\
(iii) the electron neutrino has no mixing with the state separated
with the atmospheric mass scale, or in the usual terminology the
$U_{e3}$ element vanishes at tree-level;\\
(iv) if non-vanishing, $U_{e3}$ is purely
imaginary~\cite{babu:2002dz}, to leading order.  This in turn means
that the Dirac CP-phase is maximal, a feature we refer to as
maximal CP-violation;\\
(v) to leading order the Majorana
phases~\cite{schechter:1980gr,schechter:1981gk} are constrained to be
1 or $i$~\cite{grimus:2003yn} and, although physical, do not give rise
to genuine CP-violating
effects~\cite{wolfenstein:1981rk,schechter:1981hw}.

Within SUSY theories new contributions to flavour changing processes
arise from the exchange of squarks and slepton.  In particular the
contributions are non-zero if the scalar mass matrices are
off-diagonal in the basis where the corresponding fermionic mass
matrices are diagonal (the super-CKM basis).  The experimental bounds
on flavour violating (FV) interactions in the quark sector are very
strong, whereas the bounds in the lepton sector are somewhat less
severe.  It is a general problem to achieve sufficient suppression of
the SUSY FV contributions. This is the well-known SUSY flavour problem.
A popular way to suppress the magnitude of SUSY FV is to assume that
slepton masses are universal at the Planck scale in the super-CKM
basis. In such so-called Minimal Supergravity (mSUGRA)
scenario~\cite{barbieri:1982eh,chamseddine:1982jx,hall:1983iz} RGE
running down to the electroweak scale gives naturally small calculable
off-diagonal flavour violating terms.

A necessary ingredient of the present model is that the soft SUSY
breaking terms are flavour dependent.  We should therefore be
especially worried about the strong constraints on flavour violation.
In fact, in order to get sufficient splitting of the degenerate
neutrinos, large mixings and large mass splittings in the slepton
sector are required.  In particular for smaller values of the overall
neutrino mass scale, larger off-diagonal elements of the slepton mass
matrix are necessary, in potential conflict with observation.

Our approach will therefore be to derive the possible low-energy
slepton masses and mixings by using the present knowledge of the
neutrino mass matrix. Although severely constrained by bounds on
lepton flavour violation as well as the overall neutrino mass scale, we
show that the model is indeed viable. We give the predictions for
lepton flavour violations processes, such as $\tau \to \mu \gamma$.
These are within experimental reach in the very near future.
We also note that the bounds derived here can be applied to any model
having the same tree-level form of the neutrino mass matrix as in the
$A_4$ model and using SUSY FV corrections to split the degeneracy.
Rates for lepton flavour violating in other models such as the CP
violating version of the neutrino unification model considered in
Ref.~\cite{chankowski:2000fp}, and the inverse-hierarchy model in
Ref.~\cite{Kubo:2003iw} may be treated in a similar way.
 
The plan of the paper is as follows. In Sec.~\ref{a4sec} we describe
the model, and the structure of the radiative corrections, in
Sec.~\ref{anasec} we give the numerical results for the
phenomenological FV observables and the absolute scale of neutrino
mass, and conclude in Sec.~\ref{conc}.

\section{The supersymmetric $A_4$ model}
\label{a4sec}

In this section we will give an outline of the model. For further
details we refer to the original paper in Ref.\cite{babu:2002dz} and
related work~\cite{ma:2002ge,ma:2002yp,ma:2001dn}.
As already mentioned in the introduction, the $A_4$ group is the
symmetry group of even permutations of four elements.  It has four
irreducible representations; three independent singlets, which we
denote as $\underline{1},\underline{1}' $ and $\underline{1}''$ and
one $A_4$ triplet $\underline{3}$.

The usual MSSM fields are assigned the following transformation properties 
under $A_4$
\begin{equation}
 \hat{Q}_i = (\hat{u}_i,\hat{d}_i) \:\rm{and} \: \hat{L}_i=(\hat{\nu}_i,\hat{e}_i) 
\sim \underline{3}, \,\,\,\, \,\,\hat{\phi}_{1,2}
\sim \underline{1}
\end{equation}
\begin{equation}
\hat{u}_1^c,\hat{d}_1^c,\hat{e}_1^c \sim \underline{1},
\,\,\,\,\,\,
\hat{u}_2^c,\hat{d}_2^c,\hat{e}_2^c \sim \underline{1}',
\,\,\,\,\,\,
\hat{u}_3^c,\hat{d}_3^c,\hat{e}_3^c \sim \underline{1}''  
\end{equation}
Extra SU(2) singlet heavy quark, lepton and Higgs superfields
transforming as $A_4$ triplets are added, as follows,
\begin{equation}
 \hat{U}_i, \,\, \hat{U}_i^c, \,\, \hat{D}_i, \,\, \hat{D}_i^c,  \,\,
 \hat{E}_i, \,\, \hat{E}_i^c, \,\, \hat{N}_i^c, \,\, 
 \hat{\chi}_i, \sim \underline{3}  
\end{equation}
We also assume an extra $Z_3$ symmetry under which all superfields are
singlets, except the SU(2) singlet Higgs superfield $\hat{\chi} \sim
\omega$ ($A_4$ triplet) and the SU(2) singlet superfields
$\hat{u}_i^c,\hat{d}_i^c,\hat{e}_i^c \sim \omega^2$, where
$\omega=e^{2\pi i/3}$ is the cube root of unity.

The superpotential is then given by
\begin{eqnarray}\label{superW}
 \hat{W} &=& M_U \hat{U}_i \hat{U}_i^c+f_u \hat{Q}_i\hat{U}_i^c\hat{\phi}_2  
 + h_{ijk}^u \hat{U}_i \hat{u}_j^c \hat{\chi}_k \nonumber \\
         &+& M_D \hat{D}_i \hat{D}_i^c+f_d \hat{Q}_i\hat{D}_i^c \hat{\phi}_1  
 + h_{ijk}^d \hat{D}_i \hat{d}_j^c \hat{\chi}_k \nonumber \\
         &+& M_E \hat{E}_i \hat{E}_i^c+f_e \hat{L}_i\hat{E}_i^c \hat{\phi}_1  
 + h_{ijk}^e \hat{E}_i \hat{e}_j^c \hat{\chi}_k  \\
         &+& \frac{1}{2}M_N \hat{N}_i^c\hat{N}_i^c + 
             f_N\hat{L}_i\hat{N}_i^c\hat{\phi}_2  + 
             \mu \hat{\phi}_1\hat{\phi}_2  \nonumber  \\
         &+& \frac{1}{2}M_{\chi}\hat{\chi}_i \hat{\chi}_i 
+ h_{\chi} \hat{\chi}_1\hat{\chi}_2\hat{\chi}_3 \nonumber
\end{eqnarray}

Note that the $Z_3$ symmetry is explicitly broken by the soft
supersymmetric mass term $M_\chi$. On the other hand the $A_4$
symmetry gets spontaneously broken at the high scale by the
$\vev{\chi_i}$ vev's lying along the F-flat direction given as
$\vev{\chi_1} = \vev{\chi_2} =\vev{ \chi_3} = u = -M_\chi/h_\chi$.
This solution is therefore invariant under supersymmetry, which is a
necessary requirement as we want to have low energy SUSY. In fact, the
low energy effective theory of the model is nothing but the MSSM.
Note that supersymmetry is thus only broken by TeV scale soft breaking
terms. These will also break the $A_4$ symmetry and constitute the
very source of the threshold corrections to the neutrino masses. The
electroweak symmetry is broken by the vev's of the two Higgs doublets.
As usual we define $\tan(\beta)=v_2/v_1$, where $\vev{ \phi_i^0} =
v_i$.

The charged lepton mass matrix linking $(e_i,E_i)$ to $(e_j^c,E_j^c)$
is restricted by the family symmetry to the simple form
\begin{equation}\label{clepmass}
M_{eE}= \left(\matrix{ 
 0 & 0 & 0 & f_e v_1 & 0 & 0 \cr
 0 & 0 & 0 & 0 & f_e v_1 & 0 \cr
 0 & 0 & 0 & 0 & 0 & f_e v_1 \cr
 h_1^e u & h_2^e u & h_3^e u & M_E & 0 & 0 \cr
 h_1^e u & h_2^e \omega u 
 & h_3^e \omega^2 u & 0 & M_E & 0 \cr
 h_1^e u & h_2^e \omega^2 u 
 & h_3^e \omega u & 0 & 0 & M_E \cr} \right) \;. 
\end{equation}
The $\omega$ factors arise due to the way triplets and singlets form
$A_4$-invariant combinations~\cite{ma:2002yp}. This mass matrix is
sufficiently simple to allow for an analytic diagonalization. It is of
see-saw type and the effective $3 \times 3$ low energy mass matrix,
$M_{\ell}^{\rm eff}$, can be diagonalized by $M_{\ell}^{\rm diag.}=U_L
M_{\ell}^{\rm eff} I$, where the left diagonalization matrix reads
\begin{equation}
U_{L}=\frac{1}{\sqrt{3}}\left(\matrix{ 1 & 1 & 1 \cr
1 & \omega & \omega^2 \cr
1 & \omega^2 & \omega \cr} \right)  \;.
\end{equation}
The Yukawa couplings $h_i^e$, $i=1,2,3$, are chosen such that the 
three eigenvalues of $M_{\ell}^{\rm eff}$, given by 
$m_{i}=\sqrt{3}f_e v_1 u/M_E \sqrt{1+(h_i^e u)^2/M_E^2} \, h_i^e$, 
agree with the measured masses of the electron, muon and tau leptons.
Similarly, the up-type and down-type quark mass matrices have the same
structure as in Eq(\ref{clepmass})~\cite{ma:2002yp}.  Therefore, they
will be simultaneously diagonalized, implying $V_{\rm CKM} = I$ at the
high scale.  An elegant feature of the model is that, since the quark
masses are hierarchical, the low-energy CKM mixing angles are
naturally small, as they arise due to small calculable radiative
corrections. A realistic $V_{\rm CKM}$ matrix can indeed be ascribed
to radiative corrections coming from the soft SUSY breaking scalar
quark (squark) mixing terms, starting from the tree-level identity
matrix.  This can be done obeying all experimental constraints (in
particular bounds on flavour changing processes), as already shown in
Ref.~\cite{babu:1998tm}.

Here we focus on the neutrino masses.  Rotating to the flavour basis,
where the charged leptons mass matrix is diagonal, the $6\times 6$
Majorana mass-matrix for $(\nu_e,\nu_\mu,\nu_\tau,N_1^c,N_2^c,N_3^c)$
takes the simplest (type-I) see-saw form
\begin{equation}
\left(\matrix{ 0 & U_L f_N v_2 \cr
U_L^T f_N v_2 & M_N \cr} \right) \;. 
\end{equation}
so that the effective low-energy neutrino mass matrix, is given by
\begin{equation}\label{mnutree}
 M_\nu^0 = \frac{f_N^2 v_2^2}{M_N}U_L^TU_L= \frac{f_N^2 v_2^2}{M_N}\lambda_0, 
 \qquad   \lambda_0 =  
 \left(\matrix{ 1 & 0 & 0 \cr 0 & 0 & 1 \cr 0 & 1 & 0 \cr} \right)\;.
\end{equation}
Thus the tree-level neutrino mass matrix at the $M_N$ scale has
exactly degenerate neutrinos, $m_1=m_2=m_3$, and exact maximal
atmospheric mixing.

%%%%%%%%%%%%%%%%%%%%%%%%%%%%%START RAD CORRECTIONS
%\subsection{One-loop corrections}
%%%%%%%%%%%%%%%%%%%%%%%%%%%%%%%%%%%%%%%%%%%%%%%%%%
\begin{figure}
\begin{picture}(220,100)(0,-50)
\ArrowLine(0,10)(45,10)
\ArrowLine(45,10)(120,10)
\ArrowLine(120,10)(155,10)
\ArrowLine(200,10)(155,10)
\DashCArc(80,10)(40,0,180){3}
\Vertex(155,10){3} 
\DashLine(155,10)(130,-30){3}
\DashLine(155,10)(180,-30){3}
\Text(10,17)[]{$\nu_{j}$}
\Text(80,65)[]{$\tilde\ell/ \tilde\nu$}
\Text(135,17)[]{$\nu_{k}$}
\Text(80,-3)[]{$\chi^{+}/\chi^0$}
\Text(185,17)[]{$\nu_{i}$}
\Text(10,70)[]{\Large{(a)}}
\end{picture}
\begin{picture}(220,100)(0,-50)
\ArrowLine(0,10)(45,10)
\ArrowLine(80,10)(45,10)
\ArrowLine(155,10)(80,10)
\ArrowLine(200,10)(155,10)
\DashCArc(120,10)(40,0,180){3}
\Vertex(45,10){3} 
\DashLine(45,10)(20,-30){3}
\DashLine(45,10)(70,-30){3}
\Text(10,17)[]{$\nu_{j}$}
\Text(120,65)[]{$\tilde\ell/\tilde\nu$}
\Text(65,17)[]{$\nu_{k}$}
\Text(120,-3)[]{$\chi^{+}/\chi^0$}
\Text(185,17)[]{$\nu_{i}$}
\Text(10,70)[]{\Large{(b)}}
\end{picture}
\begin{picture}(220,150)(0,-40)
\ArrowLine(0,10)(45,10)
\ArrowLine(45,10)(100,10)
\ArrowLine(100,10)(155,10)
\ArrowLine(200,10)(155,10)
\DashCArc(100,10)(55,0,180){3}
\Vertex(155,10){3}
\DashLine(155,10)(155,-30){3}
\DashLine(100,10)(100,-30){3}
\Text(10,25)[]{$\nu_{j}$}
\Text(100,85)[]{$\tilde\ell/\tilde\nu$}
\Text(70,-3)[]{$\chi^{+}/\chi^0$}
\Text(130,-3)[]{$\chi^{+}/\chi^0$}
\Text(185,25)[]{$\nu_{i}$}
\Text(10,90)[]{\Large{(c)}}
\end{picture}
\begin{picture}(220,150)(0,-40)
\ArrowLine(0,10)(45,10)
\ArrowLine(100,10)(45,10)
\ArrowLine(155,10)(100,10)
\ArrowLine(200,10)(155,10)
\DashCArc(100,10)(55,0,180){3}
\Vertex(45,10){3}
\DashLine(45,10)(45,-30){3}
\DashLine(100,10)(100,-30){3}
\Text(10,25)[]{$\nu_{j}$}
\Text(100,85)[]{$\tilde\ell/\tilde\nu$}
\Text(70,-3)[]{$\chi^+/\chi^0$}
\Text(130,-3)[]{$\chi^+/\chi^0$}
\Text(185,25)[]{$\nu_{i}$}
\Text(10,90)[]{\Large{(d)}}
\end{picture}
\caption{Feynman diagram responsible for ``wave-function'' (top) and
  ``vertex'' (bottom) radiative corrections to neutrino mass. The fat
  vertex indicates an effective dimension-5 operator obtained by
  integrating out the heavy right-handed neutrinos}
\label{loopfig}
\end{figure}
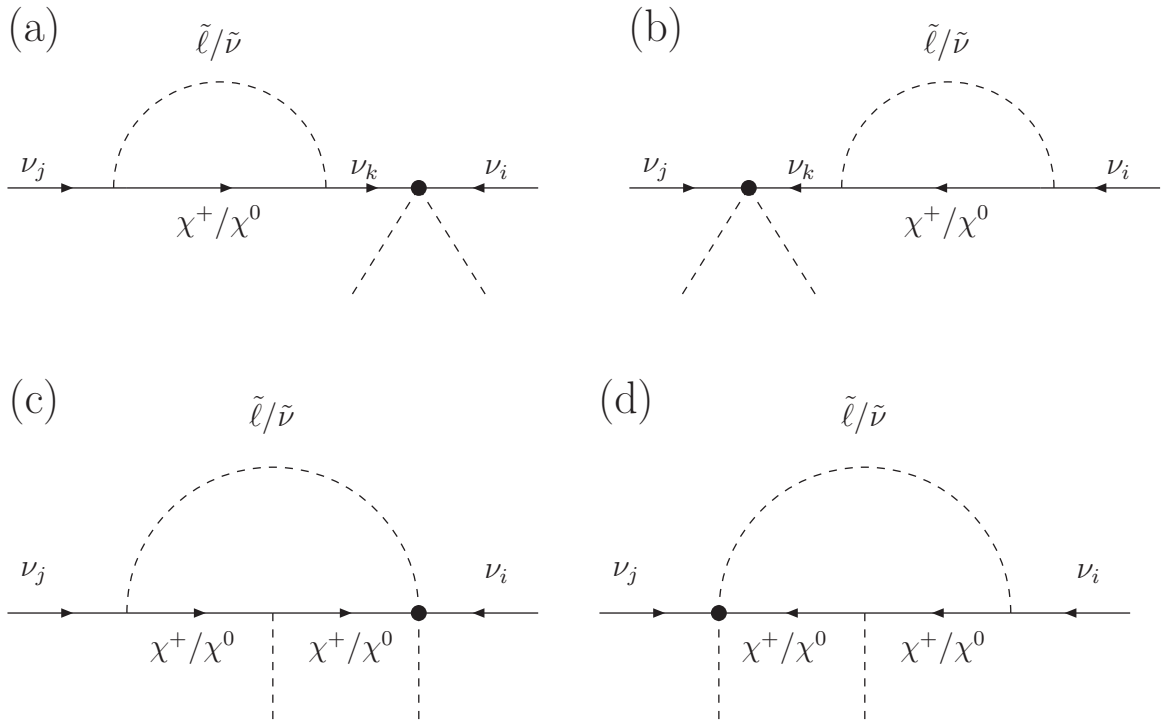

Let us turn to look at the sources of the radiative correction to the
mass matrices. 
There are in general two kinds of radiative corrections; the standard
renormalization effects arising when running from $M_N$ to the
electroweak scale, using supersymmetric renormalization group
equations (RGE's), and the low-energy threshold corrections. 

Starting with degenerate neutrinos at some high energy scale, and
using the standard minimal supergravity RGE's, it is not possible to
obtain a suitable neutrino spectrum~\cite{dighe:2000ii, casas:1999ac}.
Moreover, renormalization group evolution can not produce corrections
to the textures zeros in $M_\nu^0$, since the RGE corrections, in the
flavour basis, are proportional to the original mass
element~\cite{babu:1993qv,antusch:2001ck,chankowski:2001mx,frigerio:2002in}.
However, it is clear that small corrections to the tree-level texture
zeros are necessary in order to obtain a realistic mass matrix.

Invoking threshold corrections from flavour violating (FV) soft SUSY
breaking terms allows us to obtain both adequate neutrino mass
splittings~\cite{chun:1999vb,chankowski:2000fp} as well as mixing
angles.  We now show explicitly how a fully realistic low energy
neutrino mass matrix can be obtained when one includes the radiative
corrections coming from FV scalar lepton (slepton) interactions.
In our numerical programs we include these corrections.

The RGE effect can be approximated by
\begin{equation}
  M_{\alpha\beta}(M_S) \simeq \left[1-\frac{m_{\alpha}^2+m_{\beta}^2}
{16\pi^2 v^2\cos^2(\beta)}\log(M_N/M_S)\right] M_{\alpha\beta}(M_N) 
 \;.
\end{equation}
Let us for the moment just consider the $\tau$ Yukawa coupling. 
Then defining 
\begin{equation}
\label{raddeltatau}
  \delta_\tau \equiv
\frac{m_{\tau}^2}{8\pi^2 v^2\cos^2(\beta)}\log(M_N/M_S)
\end{equation}  
we get the values: $\delta_\tau \simeq \mathcal{O}(10^{-5})$ for
$\tan(\beta)=1$ and $\delta_\tau \simeq \mathcal{O}(10^{-3})$ for
$\tan(\beta)=15$.  Here, we have put $M_N=10^{12}$ GeV and $M_S=1000$
GeV.

In the following we calculate the radiative threshold corrections
coming from the soft SUSY breaking terms. 
At one-loop these contributions to the neutrino masses arise from the
diagrams shown in Fig.\ref{loopfig}.  For the evaluation we make some
approximations. First of all, we will not consider the full $6 \times
6$ slepton mass matrix but restrict to the $3 \times 3$ left-left part
of it.  The charged slepton mass matrix, in the super-CKM basis, may
be written as
 \begin{equation}
   M_{\tilde\ell}^2 =  \left(\matrix{ 
 M_{LL,\tilde\ell}^2  & (A - \mu \tan(\beta)) m_\ell \cr
 (A - \mu \tan(\beta)) m_\ell & M_{RR}^2
 \cr}\right) 
 \end{equation}
where each entry is a $3 \times 3$ matrix and 
\begin{equation}
  (M_{LL,\tilde\ell}^2)_{ij}= M_{L,ij}^2 
  - \frac{1}{2}(2m_{W}^2-m_{Z}^2)\cos(2\beta)\delta_{ij} 
  + m_\ell^2\delta_{ij}
\end{equation}
The $3 \times 3$ sneutrino mass matrix is given by
\begin{equation}
 (M_{LL,\tilde\nu}^2)_{ij}=
  M_{L,ij}^2+\frac{1}{2}m_{Z}^2\cos(2\beta)\delta_{ij}\;.
\end{equation}
Although there are small differences in the sneutrino and slepton
left-left mass matrices we will assume that they are identical, i.e.
$M_{LL,\tilde\ell}^2=M_{LL,\tilde\nu}^2=M_{LL}^2$.  Indeed the soft
breaking terms are expected to give the largest contribution.
Consequently the sleptons and sneutrinos have identical mixing
matrices and eigenvalues.  Let us define the mixing matrix such that
\begin{equation}
   \tilde\ell'_{\alpha} = R_{i \alpha} \tilde\ell_i \;,\qquad i=1,2,3
\end{equation}
where $\tilde\ell'_\alpha$ is the flavour eigenstate and $\tilde\ell_i$
is a mass eigenstate.  Then the mass eigenvalues can be written as
$R^{\dagger} M^2_{\rm diag.} R = M_{LL}^2$.  The contribution from the
right-sleptons in the diagrams in Fig.\ref{loopfig} are suppressed
with the Yukawa couplings squared. Hence, at least for small
$\tan(\beta)$ the approximation of only using the $3 \times 3$ slepton
mass matrix is reasonable. As we will discuss below, the solution in
agreement with data on lepton flavour violations have indeed relatively
small values of $\tan(\beta)$.

Now, as $M_{LL}$ is hermitian, it is easily realized that the structure 
of the one-loop corrections to the Majorana neutrino mass matrix can
be written as
\begin{equation}
\lambda^{\rm 1-loop} = \lambda^0 \hat{\delta} + (\hat{\delta})^T \lambda^0 
\;, \qquad \hat{\delta}=
\left(\matrix{ \delta_{ee} & \delta_{e\mu}^* & \delta_{e\tau}^* \cr
\delta_{e\mu} & \delta_{\mu\mu} & \delta_{\mu\tau}^* \cr
\delta_{e\tau} & \delta_{\mu\tau} & \delta_{\tau\tau} \cr}\right) 
\end{equation}
Therefore, the form of the neutrino mass matrix may be approximated by
\begin{equation}\label{neumass2}
M_{\nu}^{\rm 1-loop}= m_0
\left(\matrix{1+2\delta+2 \delta' & \delta'' & \delta''^* \cr 
 \delta'' & \delta & 1 + \delta - 2\delta_\tau\cr
\delta''^* & 1+\delta -2\delta_\tau & \delta \cr}\right) \;,
\end{equation}
Since the value of $\delta_0 \equiv \delta_{\mu\mu}+
\delta_{\tau\tau}-2\delta_{\mu\tau}$ does not affect the mixing
angles, it has been absorbed in $m_0$, the overall neutrino mass
scale.  Moreover, in Eq.~(\ref{neumass2}) we have defined
\begin{equation}\label{raddelta}
  \delta = 2 \delta_{\mu\tau} \;, \qquad \delta ''= \delta_{e\mu}^* + \delta_{e\tau} 
\end{equation}
\begin{equation}\label{raddeltap}
  \delta '=\delta_{ee} -\delta_{\mu\mu}/2 -\delta_{\tau\tau}/2-\delta_{\mu\tau}
\end{equation}
Note that, without loss of generality, by redefining $\nu_\mu$ and
$\nu_\tau$, one can always make the parameter $\delta$ real.  Thus,
due to the special form of $M_{\nu}^{\rm 1-loop}$ implied by the
flavour symmetry, the phase of $\delta$ can be rotated away, even
though the neutrinos are Majorana particles.

The general form of the light neutrino mixing matrix in any gauge
theory of the weak interaction containing \21 singlet leptons was
given in Ref.~\cite{schechter:1980gr}. For the case where the
isosinglets get super-heavy masses, as in the present case, it can be
approximated as a unitary matrix, $U$, which may be written as the
product of three complex rotation matrices involving three angles and
three phases, two of which are the Majorana CP-violating
phases~\cite{schechter:1980gr,schechter:1981gk}.

In the present case, due to the flavour symmetry which restricts the
form of $M_{\nu}$ as given in Eq.(\ref{neumass2}), we have that the
atmospheric mixing angle is maximal, and not affected by the radiative
corrections.  Moreover, with this parametrization the tree-level value
of the ``reactor'' angle, $s_{13}$, is zero~ \footnote{We define the
  most split neutrino mass as $m_3$ and require $m_2>m_1$, therefore
  the reactor angle is always given by $s_{13}$.}.  The model also
implies that $s_{13} \cos(\delta_{\rm CP}) = 0$, where $\delta_{\rm
  CP}$ is the Dirac CP-phase~\cite{grimus:2003yn}.
Therefore, in this model a non-zero value of $s_{13}$ implies maximal
CP-violation in the leptonic sector~\cite{babu:2002dz}.  On the other
hand, $s_{13}=0$ is equivalent to $\delta''^2$ being
real~\cite{grimus:2003yn}.  The property of maximal CP violation gives
interesting perspectives for discovery of leptonic CP-violation in
future long-baseline neutrino oscillation experiments. Finally,
Majorana CP phases affecting $\Delta L=2$ processes such as \nbb take
on CP conserving values, 0 or $\pi/2$.

In the following numerical analysis we will assume that the mass
matrix is real, in which case it can be diagonalized analytically.
The eigenvalues are given by
\begin {eqnarray}
m_1 &=& 
m_0\left( 1+2\delta+\delta'-\delta_\tau 
-\sqrt{\delta'^2+2\delta''^2+2\delta'\delta_\tau+\delta_\tau^2}
\right) \nonumber \\
m_2 &=& 
m_0\left(1+2\delta+\delta'-\delta_\tau 
+\sqrt{\delta'^2+2\delta''^2+2\delta'\delta_\tau+\delta_\tau^2}
\right) \\
m_3 &=& m_0(-1+2\delta_\tau) \nonumber
\end{eqnarray}
Hence, assuming $\delta_\tau, \delta', \delta'' \ll \delta$ the mass
squared differences can be approximated by~\footnote{Note that since
  the maximal angle has to go along with the atmospheric mass scale,
  these mass squared differences can not be swapped around.}
\begin{equation}\label{dmsqatm} 
 \Delta m^2_{\rm atm} \simeq 4 m_0^2 \delta
\end{equation}
\begin{equation}\label{dmsqsol}
 \Delta m^2_{\rm sol} \simeq 4 m_0^2  
 \sqrt{\delta'^2+2\delta''^2+2\delta'\delta_\tau+\delta_\tau^2}
\end{equation}
The solar angle is given by 
\begin{equation}\label{solangle}
   \tan^2(\theta_{\rm sol})=
\frac{2\delta''^2}{(\delta'+ \delta_\tau -
 \sqrt{\delta'^2+2\delta''^2+2\delta'\delta_\tau+\delta_\tau^2})^2}
\end{equation} 
Although generated by threshold effects, the solar angle is naturally
expected to be large, thanks to the quasi-degenerate neutrino
spectrum.
Note that if $\delta' = -\delta_\tau$ the effect of the corrections
from $\delta'$ and $\delta_\tau$ is equal to having
$\delta_0=2\delta'$ and thus amounts to an overall shift of the mass
scale. Nevertheless, in this case the solar angle would become
maximal, which is now excluded by experiments~\cite{maltoni:2003da}.

Furthermore, as $\delta'$ and $\delta_\tau$ arise from different
physics, there is no reason for this fine-tuning to take place.
Therefore, the numerical value of $\delta_\tau$ can not be much bigger
than the solar mass scale, more precisely $\delta_\tau
\stackrel{<}{\sim} 5 \times 10^{-4}$, implying that $\tan(\beta)>10$
is disfavored, as it will destroy the agreement with the solar data.
This can be seen explicitly in Fig.~\ref{brfig}.

The analytic expression for the radiative corrections to 
the neutrino masses are
\begin{eqnarray}\label{radneumass}
\delta_{\alpha \beta}^{{\rm (a)}\chi^+} &=& 
\sum_{i=1}^3 \sum_{A=1}^2|U_{A1}|^2 
\frac{g^2}{16\pi^2}B_1(m_{\chi_{A}^+}^2,m_{\tilde\ell_i}^2)
R_{i\alpha} R_{i \beta}^{\ast}  \;, \nonumber \\
\delta_{\alpha \beta}^{{\rm (a)}\chi^0} &=& 
\sum_{i=1}^3 \sum_{A=1}^4 |g N_{A2}-g' N_{A1}|^2
\frac{1}{32\pi^2}B_1(m_{\chi_{A}^0}^2,m_{\tilde\nu_i}^2)
R_{i\alpha} R_{i \beta}^{\ast}  \;, \\
\delta_{\alpha \beta}^{{\rm (c)}\chi^+} &=& 
\sum_{i=1}^3 \sum_{A=1}^2 \sum_{B=1}^2|U_{A1}V_{B2}|^2 
\frac{g^2}{4\pi^2} C_{00}(m_{\chi_{A}^+}^2,
m_{\chi_{B}^+}^2 ,m_{\tilde\ell_i}^2)
R_{i\alpha} R_{i \beta}^{\ast}  \;, \nonumber \\
\delta_{\alpha \beta}^{{\rm (c)}\chi^0} &=& 
\sum_{i=1}^3 \sum_{A=1}^4 \sum_{B=1}^4
|gN_{A2}-g'N_{A1}|^2|N_{B4}|^2 \frac{1}{8\pi^2}C_{00}(m_{\chi_{A}^0}^2,
m_{\chi_{B}^0}^2,m_{\tilde\nu_i}^2)
R_{i\alpha} R_{i \beta}^{\ast}   \;, \nonumber
\end{eqnarray}
where we have evaluated the Feynman diagrams at zero external
momentum, which is an excellent approximation as the neutrino masses
are tiny.  Here $\delta_{\alpha \beta}^{{\rm (a)}\chi^+},
\alpha,\beta=e,\mu,\tau$, is the contributions from the
chargino/slepton diagram in Fig.\ref{loopfig}(a), with analogous
notation for the other contributions.  The value of the
$\delta_{\alpha \beta}$, $\alpha,\beta=e,\mu,\tau$ in
Eq.~(\ref{radneumass}) is the sum of the four contributions given
above.  In the above formula, $U,V$ are the chargino mixing matrices
and $m_{\chi^+_A}, A=1,2$ are chargino masses.  $N$ is the neutralino
mixing matrix and $m_{\chi^0_A}, A=1,..,4$ are the neutralino masses.
The coupling constant of the $SU(2)$ gauge group is denoted $g$ and of
$U(1)$ is $g'$.  Furthermore $B_1$ and $C_{00}$ are Passarino-Veltman
functions given by
\begin{equation}
 B_1(m_0^2,m_1^2)=-\frac{1}{2}\Delta_\epsilon +\frac{1}{2} 
\ln \left( \frac{m_0^2}{\mu^2} \right)+
\frac{-3+4t-t^2-4t\ln (t)+2t^2\ln(t)}{4(t^2-1)} 
\end{equation}
where $t=m_1^2/m_0^2$ and
\begin{equation}
C_{00}(m_0^2,m_1^2,m_2^2) = 
\frac{1}{8}(3+2\Delta_{\epsilon}) -\frac{1}{4} \ln \frac{m_0^2}{\mu^2}
 +  \frac{-2r_1^2(r_2-1)\ln(r_1)+2r_2^2(r_1-1)\ln(r_2)}
{8(r_1-1)(r_2-1)(r_1-r_2)}\nonumber
\end{equation}
where $r_1=m_1^2/m_0^2$ and $r_2=m_2^2/m_0^2$. We have used
dimensional reduction, with $\epsilon=4-n$ and $n$ is the number of
space-time dimensions.  The term $\Delta_\epsilon = \frac{2}{\epsilon}
-\gamma + 4 \ln(4\pi)$, where $\gamma$ is Eulers constant, is
divergent as $\epsilon \rightarrow 0$.  However, the unitarity
conditions
\begin{equation}
  \sum_i R_{i\alpha}R_{i\beta}^* = \delta_{\alpha\beta}
\end{equation}
ensure that the infinities and the $\mu^2$ dependent terms cancel in
the corrections to the neutrino mass matrix given in
Eqs.~(\ref{raddelta})-(\ref{raddeltap}).  Therefore, the final result
does not depend on the renormalization scheme.

%%%%%%%%%%%%%%%%%%%%%%%%%%%%%%%%%%%%%

Let us shortly comment on the effect coming from the diagonalization
of the low energy charged lepton mass matrix. The neutrino mass matrix
in Eq.(\ref{radneumass}) is written in the tree-level flavour basis.
Therefore, the radiative corrections to the charged lepton masses will
result in small corrections to the mixing angles in $U$.  Indeed, for
the case of quarks, we attribute the value of $V_{\rm CKM}$ to this
type of radiative corrections.  Nonetheless, as the dominant quark
corrections originate from gluino exchange, whereas the charged
leptons only receive contributions from bino exchange, we expect that
the charged lepton mixings will be much smaller. Let us denote the
matrix which diagonalize the low energy charged lepton mass matrix by
$U_{\ell}= I + \epsilon_{\rm rad}$.  In the following we will only
consider to first order in the radiative corrections. In this case the
matrix $\epsilon_{\rm rad}$ will be anti-hermitian.  The neutrino mass
matrix $M_{\nu}^{\rm FB}$ in the flavour basis will be given by
\begin{equation}
  M_{\nu}^{\rm FB} = U_{\ell}M_{\nu}U_{\ell}^T = 
 U_{\ell}(\lambda^0 +\hat{\delta} \lambda^0 +\lambda^0 \hat{\delta})U_{\ell}^T 
 \simeq \lambda^0 +\hat{\delta} \lambda^0 +\lambda^0 \hat{\delta}^T 
          + \epsilon_{\rm rad} \lambda^0 + \lambda^0 \epsilon_{\rm rad}^T\;. 
\end{equation}
Clearly this amounts to performing the substitution $\hat{\delta} \to
\hat{\delta} + \epsilon_{\rm rad}$, but with the important difference
that $\epsilon_{\rm rad}$ is anti-hermitian whereas $\hat{\delta}$ is
hermitian.  Consequently, the atmospheric mixing angle will get small
deviations from maximal, while the CP phase will depart from its
leading order value $\pi/4$.
 
\section{Numerical Analysis and Results}
\label{anasec}

Here we test the validity of the model by performing a numerical
analysis.  We determine the allowed parameter space by searching
randomly, within certain ranges, for points in agreement with all
experimental data, considering only the radiative corrections from the
diagrams shown in Fig.~\ref{loopfig} and the effect of the RGE running
in Eq.~(\ref{raddeltatau}).  The main phenomenological restrictions
come from neutrino data and LFV constraints.

%%%%%%%%%%%%%%%%%%%%%%%%%%%%%%%%%%%%%%%%%%%%%%%%%%%%%%%%%%%%%%%%%%%%
%\subsection{CP violation in the model}

As discussed above, a non vanishing value of the reactor angle,
$s_{13}$, can only be accomplished if there is CP-violation in the
model.  From the formul{\ae}'s for the radiative corrections to the
neutrino mass matrix given in Eq.(\ref{radneumass}), it is clear that
all CP-violation originate from the slepton mixing matrix
$R_{i\alpha}$. This is true even considering the RGE effects, as the
tree-level mass matrix is real, and CP-phases cannot be generated by
RGE running.  Therefore, complex phases in the off-diagonal left-left
sleptons masses are necessary.  In Ref.\cite{bartl:2003ju} is has been
shown that these off-diagonal phases are allowed to be large.
Although, the phases in the slepton mass matrix will give
contributions to the electric dipole moment (EDM) of the electron,
they can be cancelled with contributions of other SUSY phases, such as
the phase of the $\mu$ term or phases of the gaugino masses.
Therefore, the experimental bounds on the electron EDM will not
necessarily restrict the maximum achievable magnitude of $U_{e3}$.

%%%%%%%%%%%%%%%%%%%%%%%%%%%%%%%%%%%%%%%%%%%%%%%%%%%%%%%%%%%%%%%%%%%%%%%%%
%\subsection{Lepton flavour violation}
%%%%%%%%%%%%%%%%%%%%%%%%%%%%%%%%%%%%%%%%%

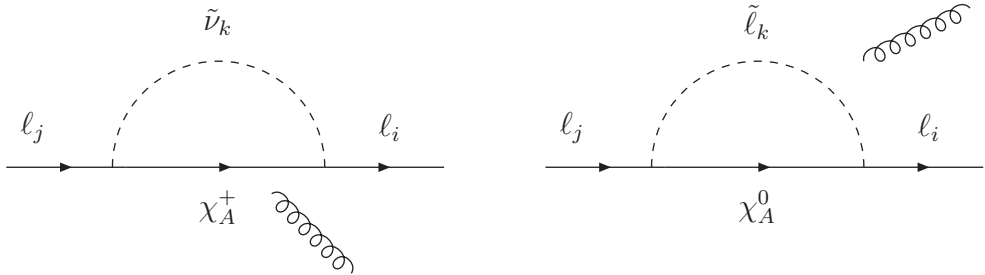
\begin{figure}
\begin{center}
\begin{picture}(200,100)(0,-30)
\ArrowLine(0,10)(45,10)
\ArrowLine(45,10)(120,10)
\ArrowLine(120,10)(165,10)
\Gluon(100,0)(130,-30){3}{6}
\DashCArc(80,10)(40,0,180){3}
\Text(10,25)[]{$\ell_{j}$}
\Text(80,65)[]{$\tilde\nu_k$}
\Text(145,25)[]{$\ell_{i}$}
\Text(80,-5)[]{$\chi^{+}_A$}
\end{picture}
\begin{picture}(200,100)(0,-30)
\ArrowLine(0,10)(45,10)
\ArrowLine(45,10)(120,10)
\ArrowLine(120,10)(165,10)
\Gluon(120,50)(160,70){3}{6}
\DashCArc(80,10)(40,0,180){3}
\Text(10,25)[]{$\ell_{j}$}
\Text(80,65)[]{$\tilde\ell_k$}
\Text(145,25)[]{$\ell_{i}$}
\Text(80,-5)[]{$\chi^{0}_A$}
\end{picture}
\end{center}
\caption{Supersymmetric contributions to the flavour violating 
charged lepton decay.}
\label{susylfv}
\end{figure}

To get sufficient suppression of LFV is a general problem in SUSY
models.  This is even more so in the $A_4$ model, since it requires
flavour violation in the slepton sector.  The strongest bounds on
lepton flavour violating processes come from $\ell_j \rightarrow \ell_i
\gamma$ and the contributions, due to exchange of SUSY particles are
shown in Fig.~\ref{susylfv}. The present bounds\footnote{Recently a
  new bound of $3.1 \times 10^{-7}$ at 90\% C.L. on $BR(\tau
  \rightarrow \mu \gamma )$ is given~\cite{abe:2003sx}.}  on these
processes~\cite{hagiwara:2002fs} are
\begin{eqnarray}
 BR(\mu \rightarrow e \gamma) &<& 1.2 \times 10^{-11} \;,\nonumber \\
 BR(\tau \rightarrow \mu \gamma) &<& 1.1 \times 10^{-6} \;, \\
 BR(\tau \rightarrow e \gamma) &<& 2.7 \times 10^{-6} \nonumber \;.
\end{eqnarray}
Explicit formulas for the SUSY contributions can be found in
Ref.\cite{hisano:1996cp}. We have used the full $6\times6$ slepton
mass matrix in order to get the slepton mixings. In so-doing we
assumed, for simplicity, that the $LR$ and $RR$ sectors of the slepton
mass matrix are flavour diagonal. In this approximation the only source
of flavour violation comes therefore from the $LL$ sector.  We have
compared our numerical results for the branching ratios against the
ones in Ref.~\cite{carvalho:2002bq,carvalho:2000xg} and found
agreement.

%%%%%%%%%%%%%%%%%%%%%%%%%%%%%%%%%%%%%%%%%%%%%%%%%%%%%%%

In total there are 10 parameters: the slepton masses and mixing
angles.  The two gaugino masses $M_1$ for the $U(1)$ gauge group and
$M_2$ for the $SU(2)$ gauge group. The value of the $\mu$ term and
$\tan(\beta)$.  For the numerical calculations we take all SUSY masses
in the range 100 GeV to 1000 GeV. If SUSY masses are much larger
supersymmetry will no longer solve the hierarchy problem, which is
indeed one of the strongest arguments for SUSY.  The results that we
present in the following are quite naturally dependent on the upper
cut on the SUSY masses.  If larger masses are admitted, the allowed
parameter space will be larger.  Furthermore, all parameters are taken
to be real and therefore $s_{13}=0$ is obtained. Hence, the reactor
bound is automatically satisfied. The neutrino parameters are taken
within 3$\sigma$ ranges allowed by the most recent solar, atmospheric,
reactor and accelerator data, taken from Ref.~\cite{maltoni:2003da}.
The most relevant parameters in our analysis are the solar angle and
mass squared difference, as well as the atmospheric mass squared
difference. These two mass splittings may potentially conflict with
the overall mass scale $m_0$ for the degenerate neutrinos, fixed at
the large scale where the flavour symmetry holds. The absolute
neutrino mass scale is constrained by cosmology and by
$(\beta\beta)_{0\nu}$ experiments.  Using the recent data from
WMAP~\cite{bennett:2003bz} and 2df galaxy survey~\cite{elgaroy:2002bi}
a bound on the sum of neutrino masses in the range $0.7-1.0$ eV (95\%
CL.)  has been claimed~\cite{Hannestad:2003xv,spergel:2003cb}.
However, a more recent re-analysis dropping prior assumptions gives a
less stringent bound of 1.8 eV which leads to~\cite{Tegmark:2003ud}
\begin{equation}
\label{m0bound}
 m_0 < 0.6 \, {\rm eV} \;.
\end{equation}
Similarly, from neutrinoless double beta decay an upper bound is
obtained, less strict, considering the nuclear matrix element
uncertainties (there also exists claims in favor of degenerate
neutrino~\cite{allen:2003pt,klapdor-kleingrothaus:2001ke}. See,
however, Ref.~\cite{Aalseth:2002dt}).

Here we take the conservative upper limit of 0.6 eV on the magnitude
of the Majorana neutrino mass $m_0$. In order to obtain the measured
atmospheric mass squared difference, approximately given by
Eq.(\ref{dmsqatm}), a minimum value for the $\delta$ parameter,
defined in Eq.(\ref{raddelta}), is needed.  Similarly to arrive at the
right solar mass squared difference, roughly given by
Eq.(\ref{dmsqsol}), the values of $\delta'$ and $\delta''$, defined in
Eqs.(\ref{raddelta})-(\ref{raddeltap}), should be around
$10^{-5}-10^{-4}$.  Furthermore, it is clear from Eq.(\ref{solangle}),
that $\delta '' \sim \delta '$ is necessary in order to obtain a large
solar mixing. The numerical study gives the bound $|\delta'/\delta''|
> 0.1$.
  
\begin{figure}
 \centering
  \includegraphics[width=0.55\textwidth]{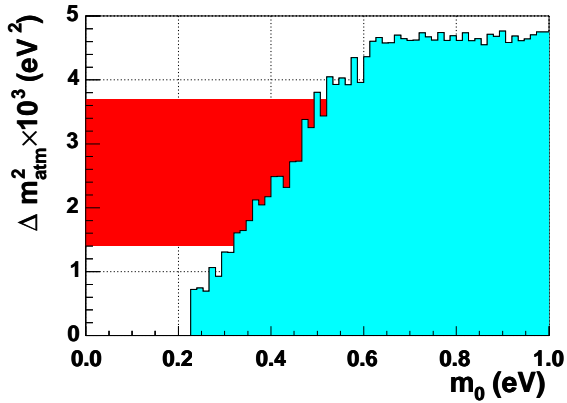}
\caption{The light shaded histogram shows the maximum possible value 
  of the atmospheric mass squared difference as a function of $m_0$.
  The dark shaded region is the current 3$\sigma$ allowed region for
  $\Dma$ from \cite{maltoni:2003da}.}
\label{mzerofig}
\end{figure}

It is non-trivial to obtain large enough values for the
$\delta,\delta'$ and $\delta''$ parameters.  To produce a large value
of $\delta$, large mass splittings as well as large mixing in the
$\tilde{\mu}-\tilde{\tau}$ sector are needed.  The upper bound on the
mass gaps will give a maximum value for $\delta$ which, through the
relation to $\Dma$, imposes a minimum value for $m_0$.  In
Fig.\ref{mzerofig} the maximum achievable value of $\Dma$ is plotted
against the value of $m_0$.  A conservative lower bound
\begin{equation}
 m_0 \geq 0.3 \, {\rm eV} \;
\end{equation}
is derived. This is very close to the present limit from experiments,
which we take as Eq.~\ref{m0bound}.

The spectrum for the charged sleptons, which is taken to be the same
as that of the sneutrinos, fall in two different classes.  One group,
which we will call the normal hierarchy, have two low mass sleptons
and the third mass rather large (around 800 GeV). The second group,
denoted the inverted hierarchy case, has two rather large masses in
the neighborhood of 700 GeV and the lightest mass typically at
150~GeV. In either case, at least one slepton mass lies below $\sim
200$ GeV which is detectable, for example, at the LHC.  Most points
fall into the case of normal hierarchy, which as a matter of fact
often corresponds to a normal hierarchy for the neutrinos as well
($\delta<0$). 

\begin{figure}
\begin{center}
   \includegraphics[height=5cm,width=0.4\textwidth]{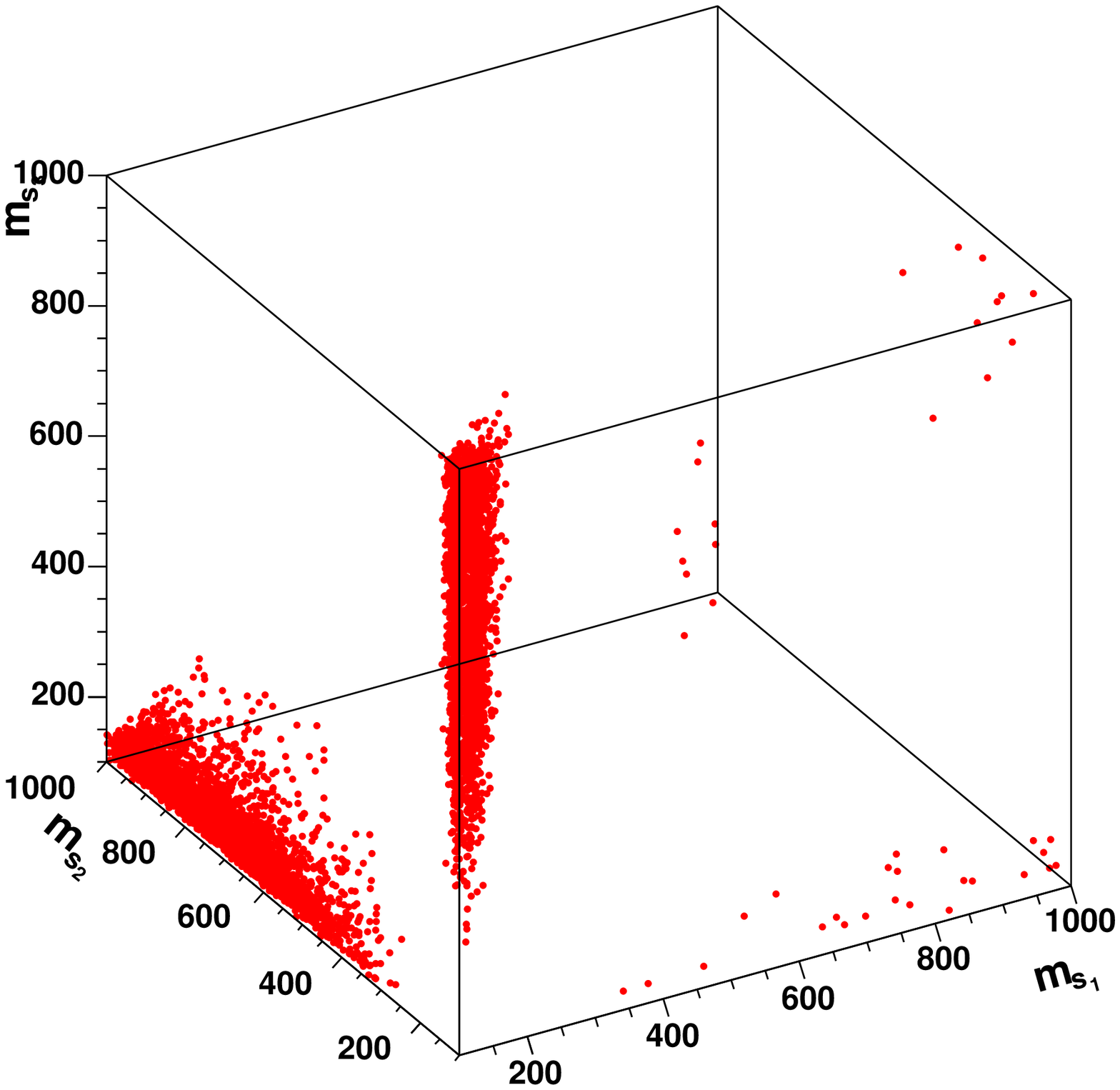}
    \includegraphics[height=5cm,width=0.4\textwidth]{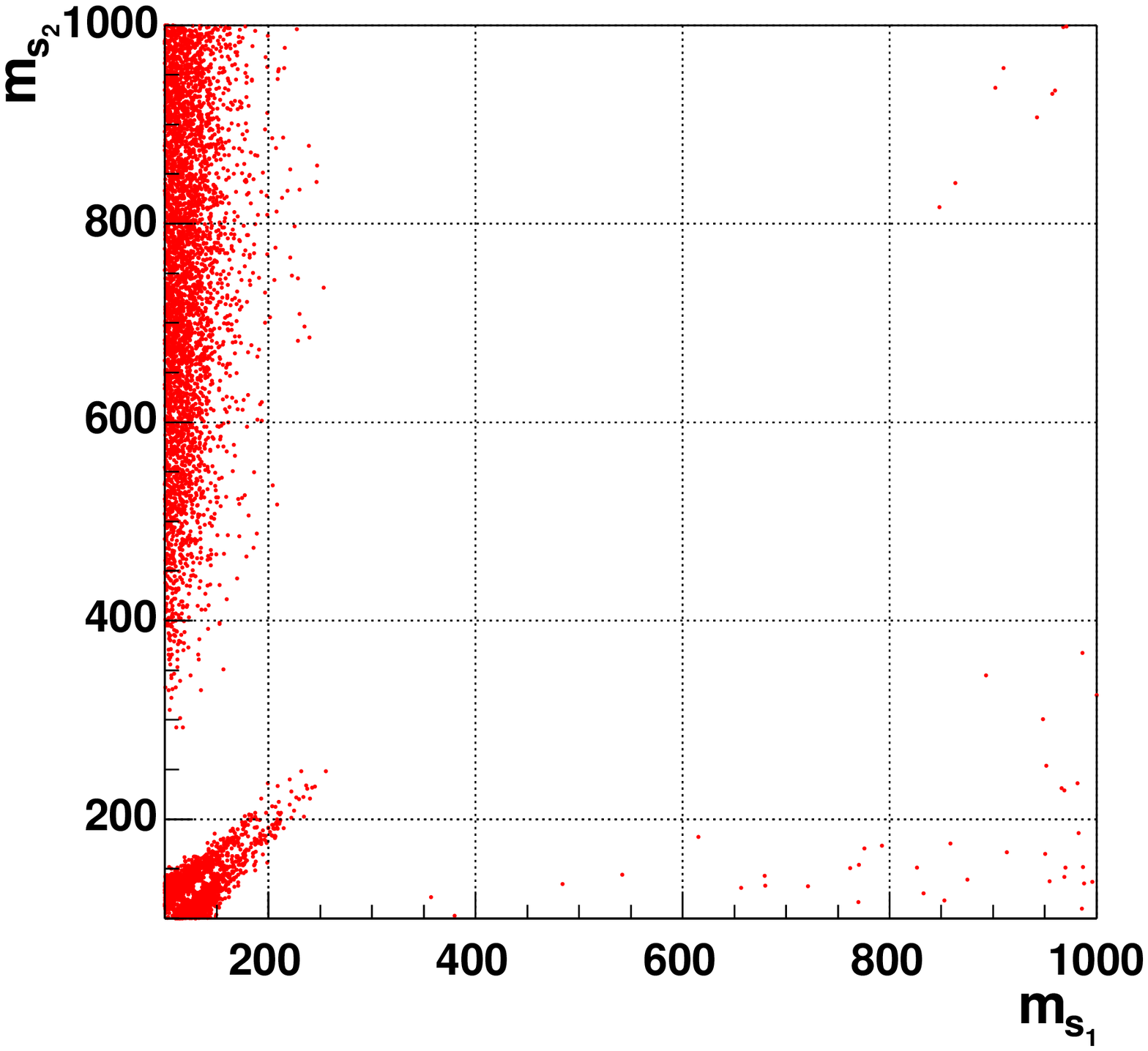} \\
    \includegraphics[height=5cm,width=0.4\textwidth]{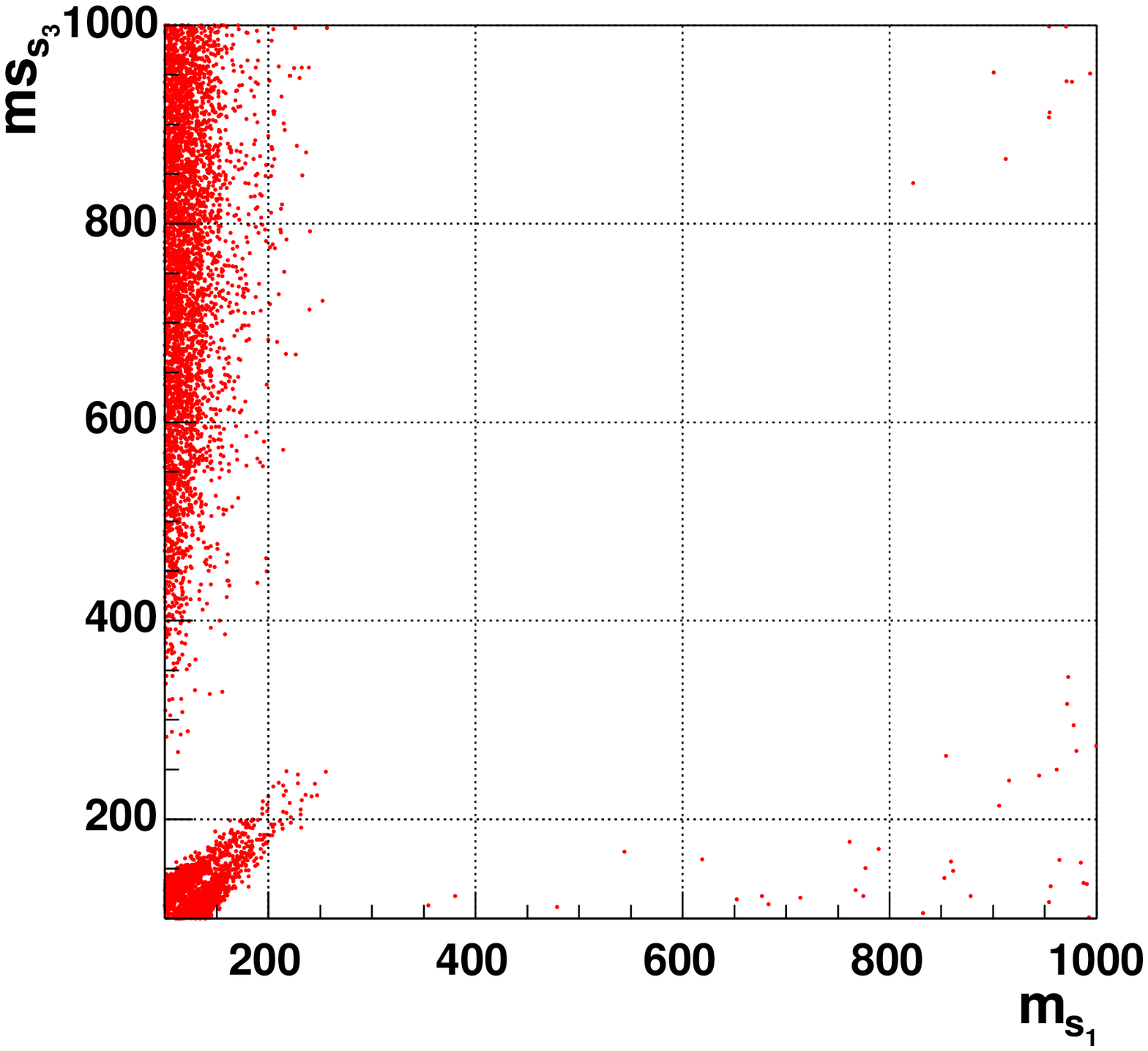}
    \includegraphics[height=5cm,width=0.4\textwidth]{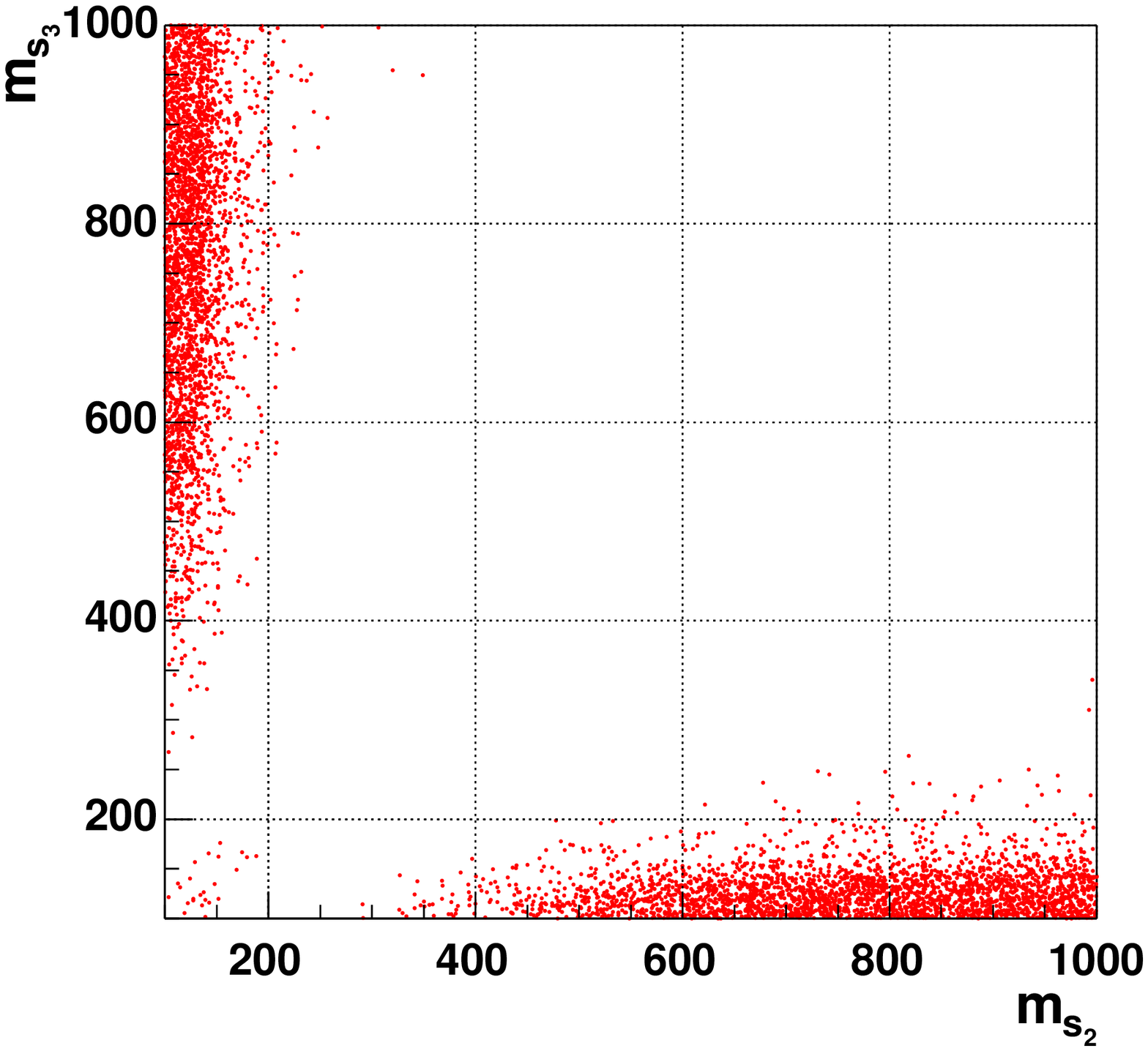}
\end{center}
\caption{The slepton and sneutrino masses for the normal hierarchy
  case, $\delta<0$}
\label{massfig}
\end{figure}

\begin{figure}
\begin{center}
    \includegraphics[height=5cm,width=0.4\textwidth]{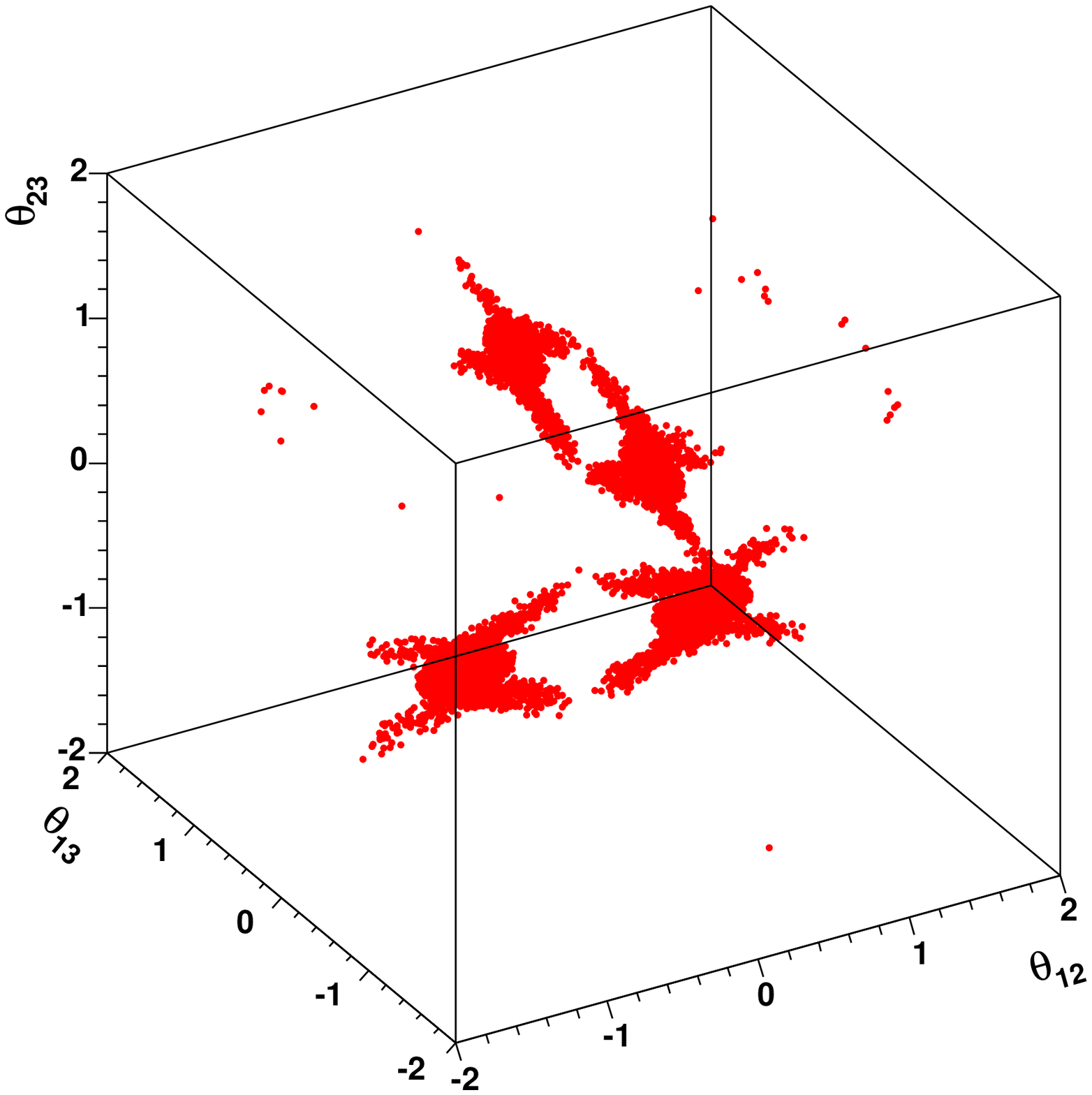}
    \includegraphics[height=5cm,width=0.4\textwidth]{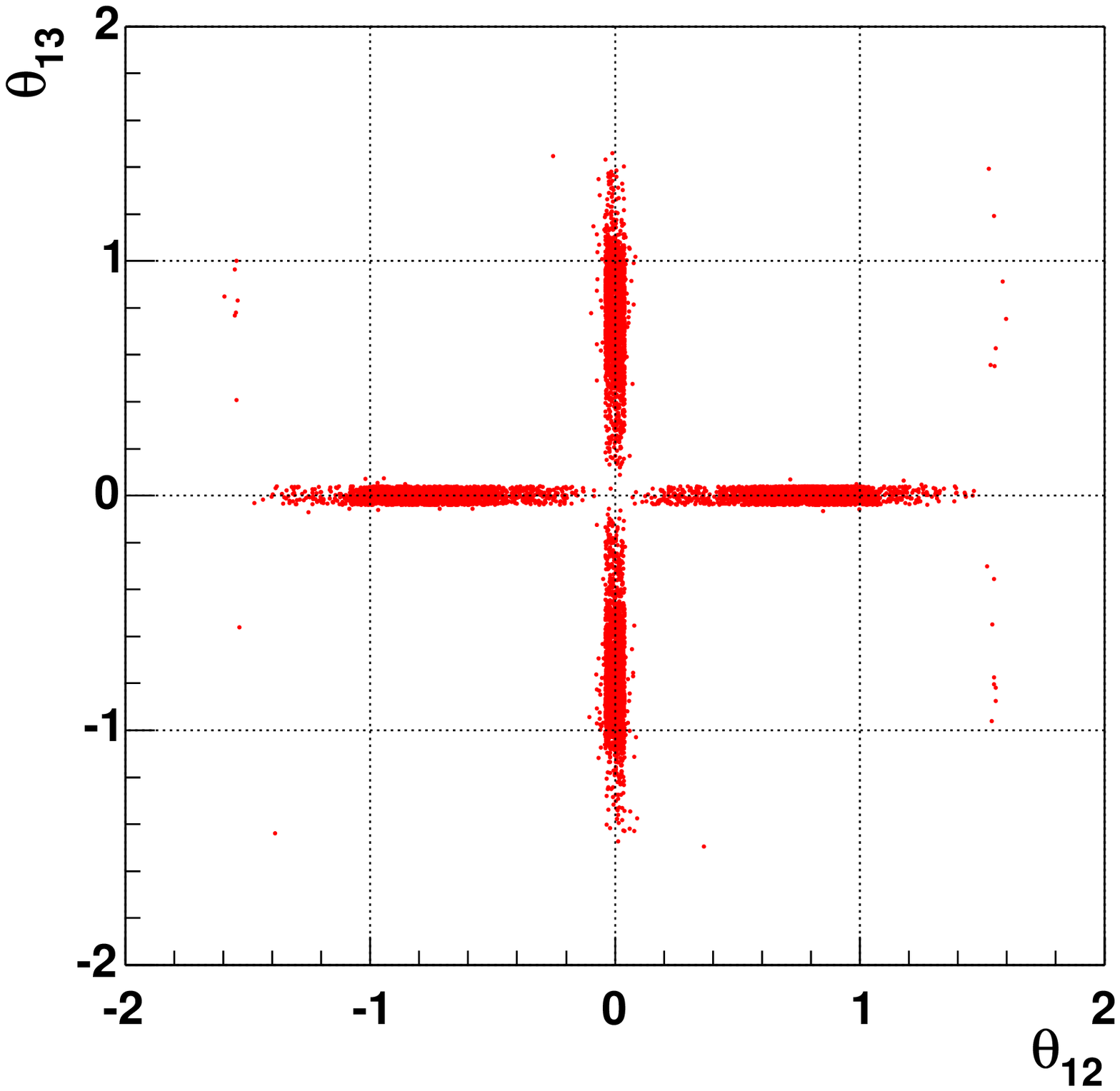} \\
    \includegraphics[height=5cm,width=0.4\textwidth]{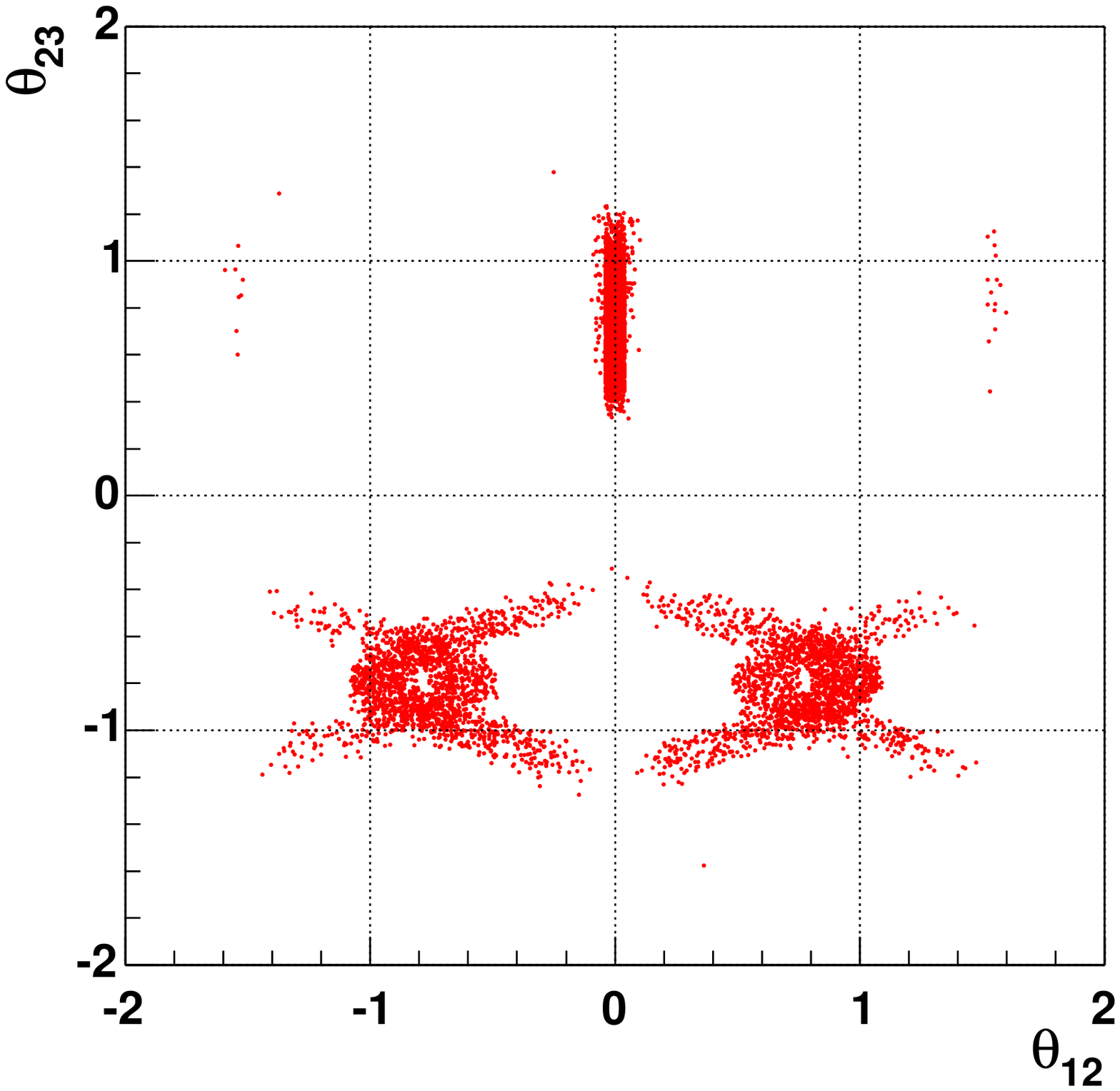}
    \includegraphics[height=5cm,width=0.4\textwidth]{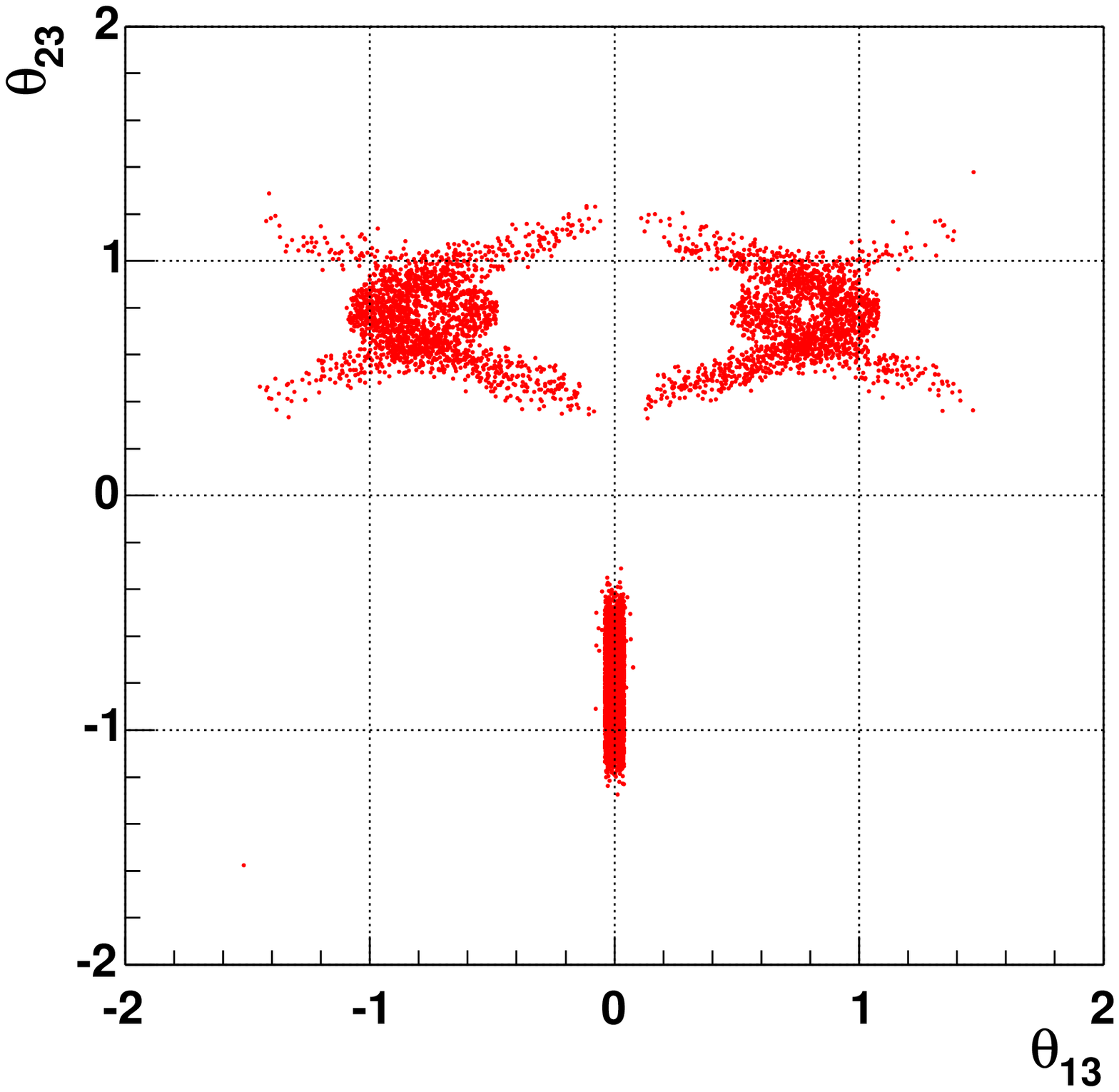}
\end{center}
\caption{The slepton and sneutrino mixing angles for the normal hierarchy
  case, $\delta<0$}
\label{mixfig}
\end{figure}

\begin{figure}
\begin{center}
\begin{picture}(360,180)(0,0)
\CBox(0,160)(1,175){Black}{Blue}
\CBox(1,160)(80,175){Black}{Yellow}
\CBox(80,160)(160,175){Black}{Red}
\CBox(0,25)(64,40){Black}{Blue}
\CBox(64,25)(112,40){Black}{Yellow}
\CBox(112,25)(160,40){Black}{Red}
\CBox(0,5)(96,20){Black}{Blue}
\Text(48,12)[c]{\textcolor{white}{$\tilde{e}$}}
\CBox(96,5)(128,20){Black}{Yellow}
\Text(112,12)[c]{\textcolor{black}{$\tilde{\mu}$}}
\CBox(128,5)(160,20){Black}{Red}
\Text(144,12)[c]{\textcolor{black}{$\tilde{\tau}$}}
\CBox(200,5)(201,20){Black}{Blue}
\CBox(201,5)(280,20){Black}{Yellow}
\CBox(280,5)(360,20){Black}{Red}
\CBox(200,160)(296,175){Black}{Blue}
\Text(248,167)[c]{\textcolor{white}{$\tilde{e}$}}
\CBox(296,160)(328,175){Black}{Yellow}
\Text(312,167)[c]{\textcolor{black}{$\tilde{\mu}$}}
\CBox(328,160)(360,175){Black}{Red}
\Text(345,167)[c]{\textcolor{black}{$\tilde{\tau}$}}
\CBox(200,140)(264,155){Black}{Blue}
\CBox(264,140)(312,155){Black}{Yellow}
\CBox(312,140)(360,155){Black}{Red}
\end{picture}
\end{center}
\caption{The rough form of the slepton and sneutrino spectrum 
in the case of normal hierarchy (left)
or inverted hierarchy (right).}
\label{snuspecfig}
\end{figure}
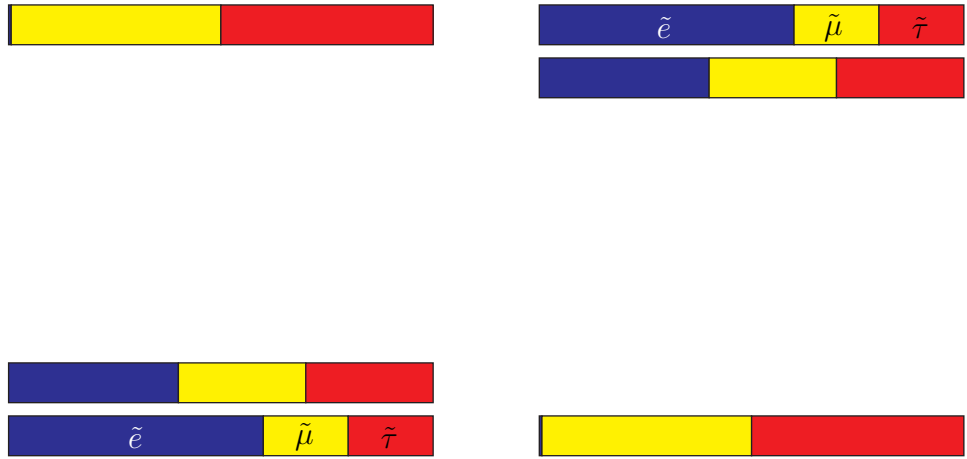

In Fig.\ref{massfig} we display the slepton masses, and in
Fig.\ref{mixfig} the mixing angles are shown for the normal hierarchy.
It is clearly seen that the spectrum contains two large mixing angles
and one small mixing angle, needed to suppress the decay $\mu \to e
\gamma$.  Also the degeneracy of two of the sleptons helps to minimize
the LFV.  As a rule of thumb there is at least one pair of sleptons
with a mass splitting of less than 40 GeV. The rough spectrum needed
is schematized in Fig.\ref{snuspecfig}.  
Although there is room for substantial deviations from the spectrum
shown, the similarity with the neutrino spectrum is quite striking:
the large mixings in the slepton sector are rather correlated to the
large mixings in the neutrino sector.  We also obtain a lower bound on
the value of the $\mu$ parameter around 500 GeV, although in the case
of inverted hierarchical slepton spetrum there are a few points with
$\mu \sim 200$ GeV. Thus in most cases the second chargino, which is
almost pure Higgsino is rather heavy.

As there is at least one low--mass slepton present in the model, one
could suspect that a large contribution to the anomalous magnetic
moment of the muon will result. We have explicitely calculated the
magnitude, as in Ref.~\cite{moroi:1996yh,martin:2001st}.  The rough
order of magnitude is $10 \times 10^{-10}$, which is too small to
explain the BNL result~\cite{Bennett:2002jb}.  As is well-known the
contribution to $g-2$ has the same sign as the $\mu$-term, thereby
disfavoring negative values for the $\mu$ parameter. 

\begin{figure}
 \includegraphics[height=5cm,width=0.4\textwidth]{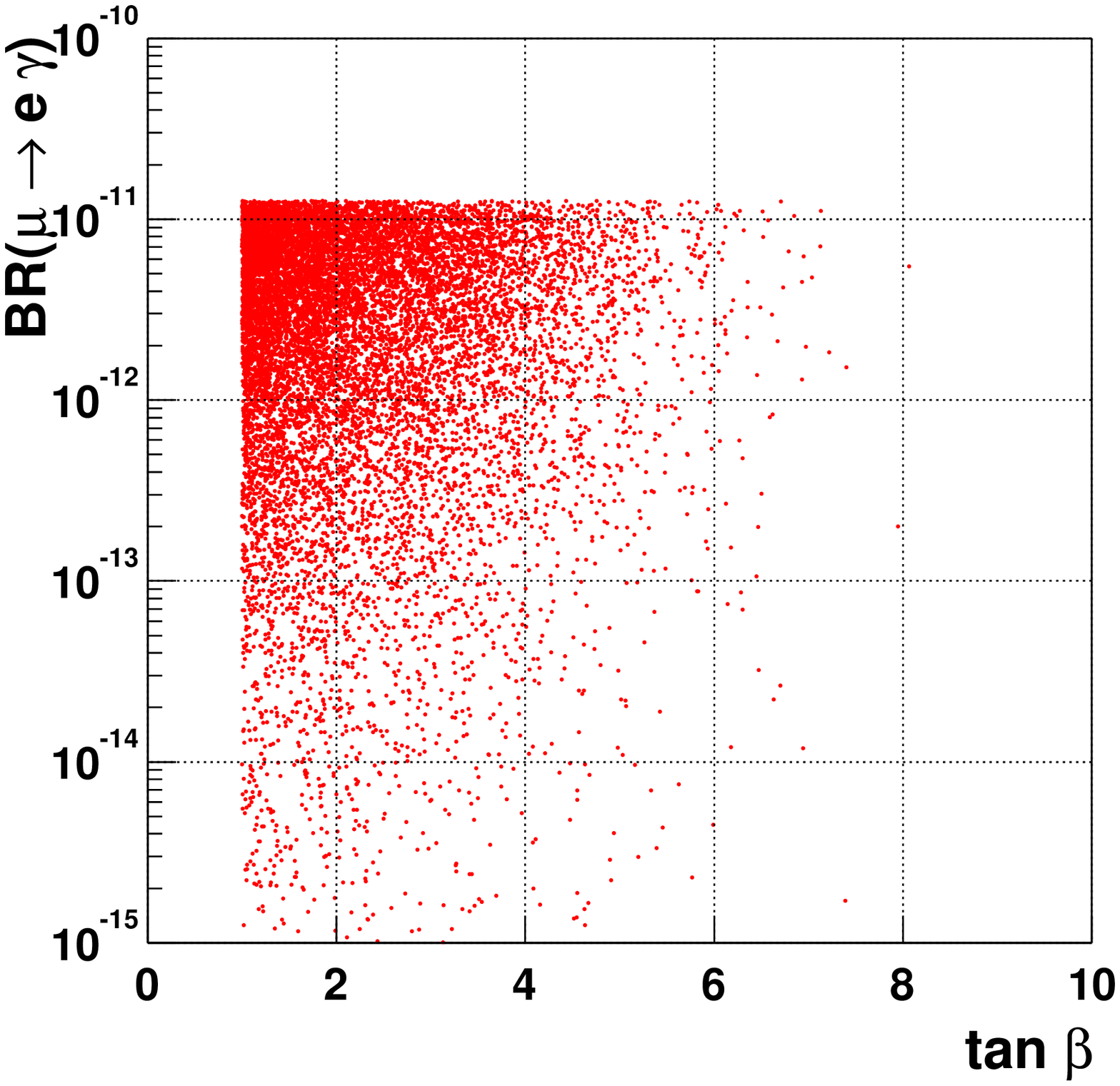}
  \includegraphics[height=5cm,width=0.4\textwidth]{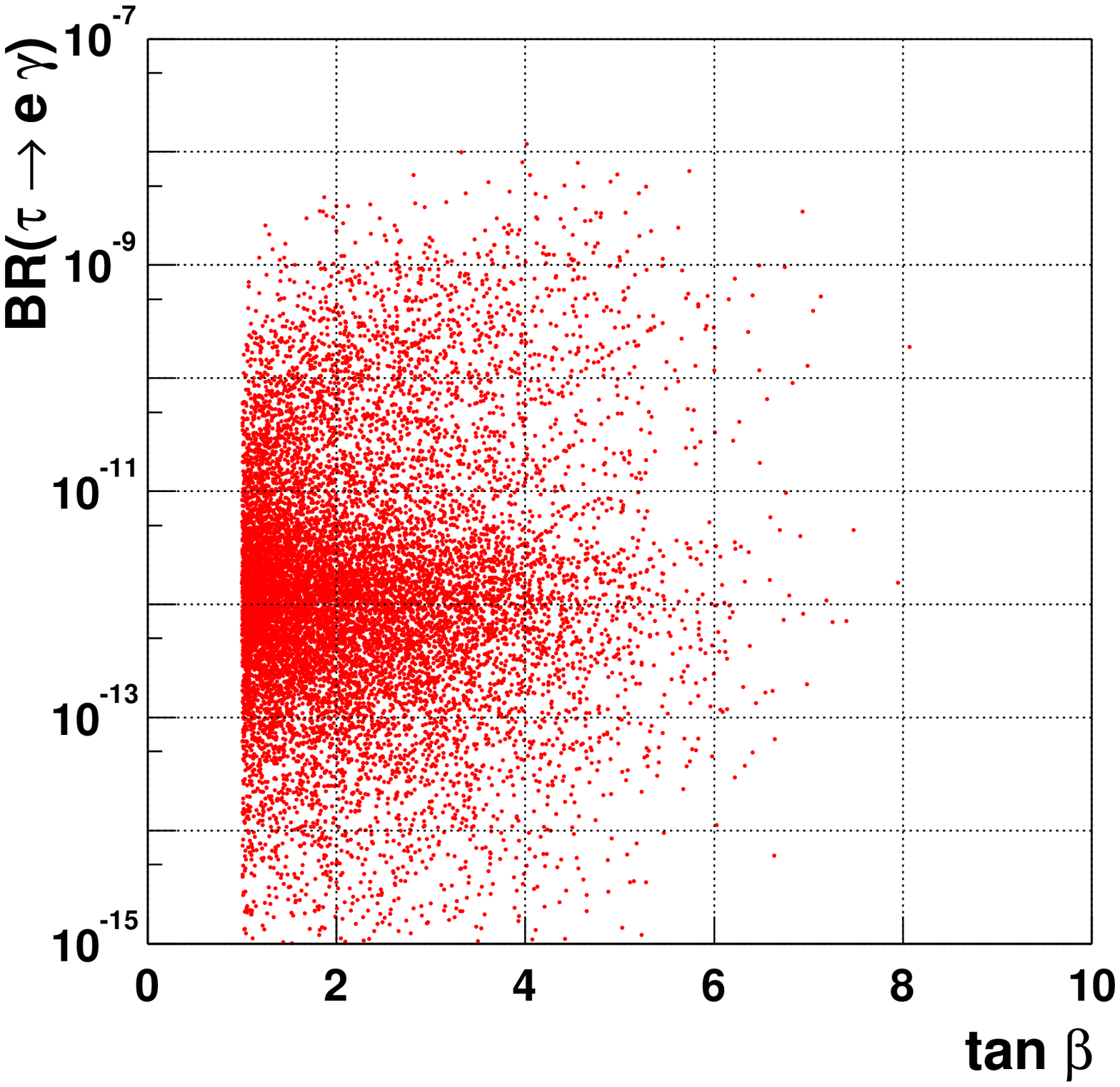}\\
  \includegraphics[height=5cm,width=0.4\textwidth]{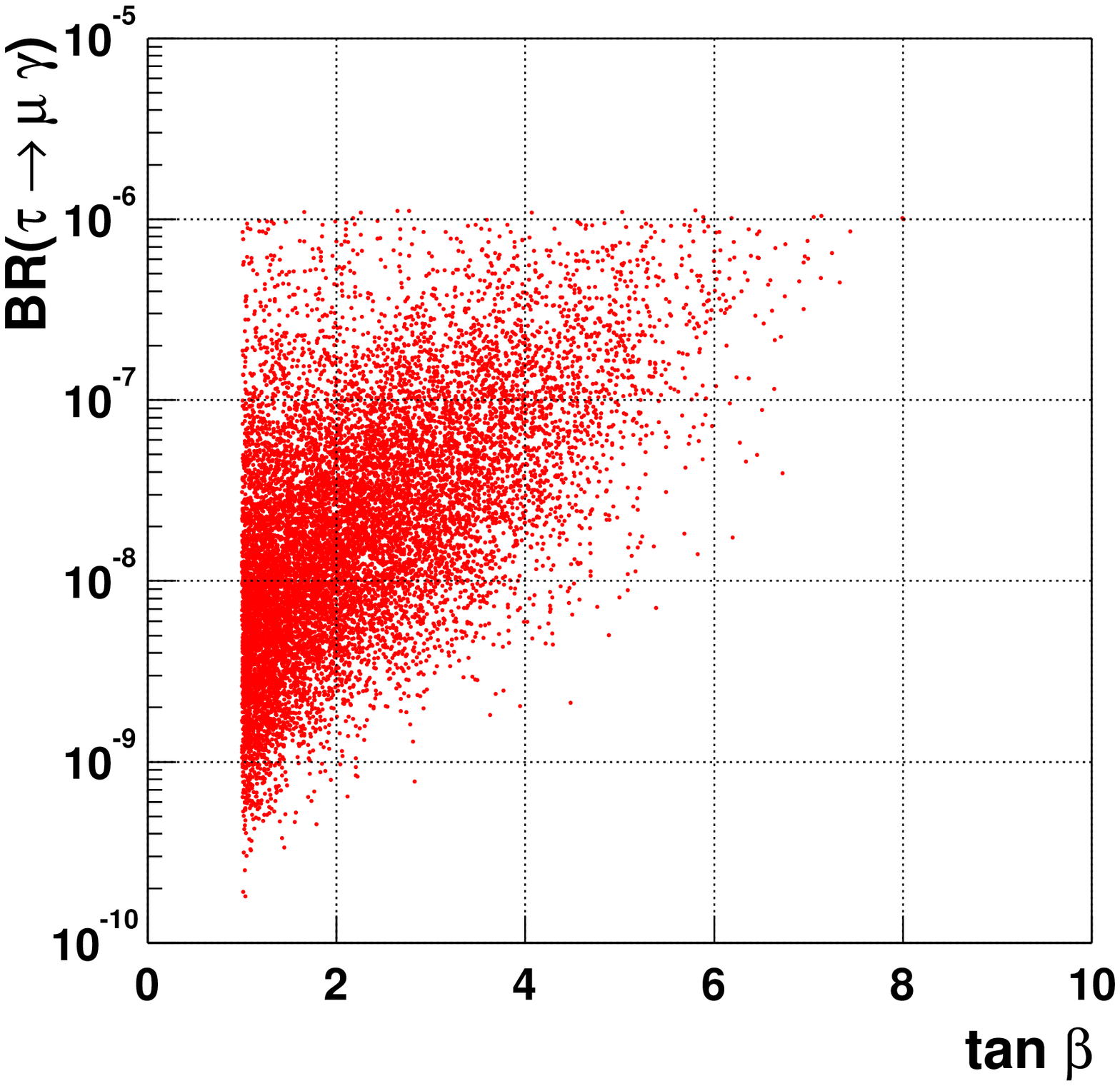}
\caption{The branching ratios for the processes 
$\ell_i \rightarrow \ell_j \gamma$ as a 
function of $\tan(\beta)$.}
\label{brfig}
\end{figure}

An important outcome of our study is the prediction for the LFV
radiative charged lepton decays~\footnote{Note that, due to the
  presence of isosinglet charged leptons, this model implies the
  existence of tree-level LFV decays such as $\mu \to 3 e$.  However,
  due to the large value of $M_E$, close to the $A_4$ scale, their
  expected magnitude would be too small to be phenomenologically
  relevant.  } $\ell_i \to \ell_j \gamma$. As seen in Fig.~\ref{brfig}
a lower bound of $10^{-9}$ for BR$(\tau \to \mu \gamma)$ is found.
This is within reach of the future BaBar and Belle
search~\cite{roney:2002ba}. The model also leads to sizeable rates for
muon-electron conversion. We find that BR$(\mu \to e \gamma)$ is
constrained to be larger than about $10^{-15}$ and therefore stands
good changes of being observed at future LFV searches, such as those
taking place at PSI.  As already noted, Fig.~\ref{brfig} indicates the
existence of an upper bound on the value of $\tan(\beta)$ in this
model. Taking the new constraint given in Ref.~\cite{abe:2003sx}, this
upper limit would be $\tan(\beta)\lsim 7$.

In conclusion our analysis shows that, although we can obtain a
realistic model in a fully consistent range of the SUSY parameter
space, the model is rather strongly restricted and will be tested in
the near future in a crucial way.

\newpage
%%%%%%%%%%%%%%%%%%%%%%%%%%%%%%%%%%%%%%%%%%%%%%%%%%%%%%%%%%%%%%%%%%%%%%%%
\section{Conclusion}
\label{conc}

The $A_4$ model that we have studied provides a complete picture of
the flavour structure, especially it offers a common mechanism to
obtain viable quark and lepton mixing matrices.  The flavour dependence
of the soft supersymmetry breaking terms acts as the source of the
radiative corrections to the fermion masses. It splits the degeneracy
of the neutrino masses as well as the alignment of the quark masses.
We have shown that starting from a three-fold degeneracy at a high
energy scale, it is possible to obtain a mass matrix in complete
agreement with all current neutrino data.  Within the model the lepton
and slepton mixings are intimately related, with one slepton mass
lying below 200 GeV. The flavour composition of this state ensures
that it will be detectable at future collider experiments, such as the
LHC.
 
The radiative corrections restrict the form of the neutrino mass
matrix imply,
\begin{itemize} 
\item Maximal atmospheric mixing. 
\item Maximal leptonic CP violation (unless $U_{e3}=0$).
\end{itemize} 

Note that the maximality of leptonic CP violation is a feature of the
leading order approximation and may acquire sizeable corrections.

The absolute Majorana neutrino mass scale for the quasi-degenerate
neutrinos is shown to be larger than 0.3 eV and is bounded from above by
cosmology as in Eq.~(\ref{m0bound}), and therefore lies in the range
of sensitivity of upcoming searches for neutrinoless double beta
decay~\cite{klapdor-kleingrothaus:1999hk}, and tritium beta
decay~\cite{osipowicz:2001sq}.
 
We have also shown how the model is fully consistent with current data
on lepton flavour violation. The predictions of lepton flavour violating
charged lepton decays lie in a range accessible to future tests.  We
find, for example, that the BR$(\mu \to e \gamma)$ lies close to the
current experimental limits, although parameters can easily be chosen
so that the bound is obeyed.  On the other hand we find a lower bound
for the $\tau \to \mu \gamma$ decay branching ratio, BR$(\tau \to \mu
\gamma)> 10^{-9}$.

Let us also mention the fact that the study we have performed is not
only restricted to the specific $A_4$ model presented in section
(\ref{a4sec}).  Any model with the neutrino mass matrix given by
$\lambda_0$ at some high energy scale and with supersymmetric flavour
changing radiative corrections will have the same constraints as
presented in this work.

\section{Acknowledgements}

We thank K. S. Babu and M. Gomez for useful discussions.  This work
was supported by Spanish grant BFM2002-00345, by European RTN network
HPRN-CT-2000-00148, by European Science Foundation network grant
N.~86. M.~H. acknowledges support from the Ramon y Cajal programme.

\end{document}